\documentclass[journal,10pt]{IEEEtran}

\IEEEoverridecommandlockouts
\usepackage{cite}
\usepackage{amsmath,amssymb,amsfonts}
\usepackage{algorithmic}
\usepackage{graphicx}
\usepackage{textcomp}
\usepackage{xcolor}
\def\BibTeX{{\rm B\kern-.05em{\sc i\kern-.025em b}\kern-.08em
    T\kern-.1667em\lower.7ex\hbox{E}\kern-.125emX}}
\usepackage{graphicx}
\usepackage{subcaption}
\graphicspath{ {images/} }
\usepackage[utf8]{inputenc}
\usepackage{fancyhdr}
\usepackage{dirtytalk}
\usepackage{tabularx}
\PassOptionsToPackage{hyphens}{url}\usepackage{hyperref}
\usepackage[ruled,lined,linesnumbered]{algorithm2e}

\usepackage{enumitem,kantlipsum}

\usepackage{graphicx}
\graphicspath{ {images/} }
\usepackage[utf8]{inputenc}
\usepackage{fancyhdr}
\usepackage{dirtytalk}
\usepackage{tikz}
\usetikzlibrary{shapes.geometric, arrows}
\usepackage{pgfplots}
\pgfplotsset{width=8cm,compat=1.16}
\usepgfplotslibrary{external}
\usepackage{mathtools}

\tikzstyle{startstop} = [rectangle, rounded corners, minimum width=2cm, minimum height=0.5cm,text centered, draw=black, fill=red!30]
\tikzstyle{process} = [rectangle, minimum width=2cm, minimum height=0.5cm, text centered, draw=black, fill=orange!30, align=left]
\tikzstyle{decision} = [diamond, minimum width=1.0cm, minimum height=0.4cm, text centered, draw=black, fill=green!30]
\tikzstyle{arrow} = [thick,->,>=stealth]

\newcolumntype{L}{>{$}l<{$}}

\usepackage[labelformat=simple]{subcaption}

\begin{document}


\title{VMGuard: Reputation-Based Incentive Mechanism for Poisoning Attack Detection in Vehicular Metaverse}


\author{Ismail Lotfi, Marwa Qaraqe, Ali Ghrayeb and Dusit Niyato

\thanks{
\textit{(Corresponding author: Marwa Qaraqe)}

This work was made possible by AICC03-0324-200005 from the Qatar National Research Fund (a member of Qatar Foundation). The findings herein reflect the work, and are solely the responsibility, of the authors. 

Ismail Lotfi, Marwa Qaraqe and Ali Ghrayeb are with the Division of Information and Computing Technology, College of Science and Engineering,
Hamad Bin Khalifa University, Qatar Foundation, Doha, Qatar (e-mail: ismail003@e.ntu.edu.sg, mqaraqe@hbku.edu.qa, aghrayeb@hbku.edu.qa).

Dusit Niyato is with the School of Computer Science and Engineering, Nanyang Technological University, Singapore (e-mail: dniyato@ntu.edu.sg).
}
}

\maketitle
\begin{abstract}


The vehicular Metaverse represents an emerging paradigm that merges vehicular communications with virtual environments, integrating real-world data to enhance in-vehicle services. However, this integration faces critical security challenges, particularly in the data collection layer where malicious sensing IoT (SIoT) devices can compromise service quality through data poisoning attacks.
The security aspects of the Metaverse services should be well addressed both when creating the digital twins of the physical systems and when delivering the virtual service to the vehicular Metaverse users (VMUs). 
This paper introduces vehicular Metaverse guard (VMGuard), a novel four-layer security framework that protects vehicular Metaverse systems from data poisoning attacks. 
Specifically, when the virtual service providers (VSPs) collect data about physical environment through SIoT devices in the field, the delivered content might be tampered. Malicious SIoT devices with moral hazard might have private incentives to provide poisoned data to the VSP to degrade the service quality (QoS) and user experience (QoE) of the VMUs.
The proposed framework implements a reputation-based incentive mechanism that leverages user feedback and subjective logic modeling to assess the trustworthiness of participating SIoT devices. 
More precisely, the framework entails the use of reputation scores assigned to participating SIoT devices based on their historical engagements with the VSPs. These scores are calculated from feedback provided by the VMUs to the VSPs regarding the content they receive and are managed utilizing a subjective logic model.
Ultimately, we validate our proposed model using comprehensive simulations. Our key findings indicate that our mechanism effectively prevents the initiation of poisoning attacks by malicious SIoT devices. Additionally, our system ensures that reliable SIoT devices, previously missclassified, are not barred from participating in future rounds of the market.
This work provides a crucial security solution for the emerging vehicular Metaverse, enabling trustworthy data collection and reliable service delivery.
\end{abstract}


\section{Introduction}
\subsection{Background and Motivation}
The Metaverse, seen as the Internet's upcoming evolution, has rapidly expanded across various domains like remote work-spaces, vehicular systems, and online banking~\cite{Minrui_COMST_2023, ISMAIL_JSAC_2023_semantic}. Technologies such as virtual reality (VR), augmented reality (AR), and artificial intelligence (AI) play pivotal roles in its development. The introduction of Meta's Quest 3 brings mixed reality (XR), another crucial element for the deployment of the Metaverse, closer to realization than ever\footnote{https://www.meta.com/quest/quest-3/}. Anticipating a mobile-centric user base, especially in contexts like the vehicular Metaverse, the design of next generation mobile networks become crucial to adequately support diverse Metaverse services.

The initial step in exploring the potential of the vehicular Metaverse involves creating a digital replica of the physical world. To achieve this, the virtual service provider (VSP) engages sensing IoT (SIoT) devices such as unmanned aerial vehicles (UAVs), smart cars, and Closed-circuit television (CCTV) cameras to gather environmental data, mainly comprising images and video scenes. However, this influx of data leads rapidly to network congestion. 
Additionally, incentive mechanisms to encourage data sharing by SIoT devices needs to be well designed.
To tackle these challenges, prior studies propose semantic communication to alleviate data congestion and enhance data freshness~\cite{Ismail_FNWF_2022, ISMAIL_JSAC_2023_semantic, Jiacheng_TWC_2022}. Semantic communication involves using machine learning algorithms to extract meaningful information from raw data then transmitting only this extracted data to the VSP. 
Incentivizing SIoT devices involves adapting auction and contract theories, ensuring individual rationality (IR) and incentive compatibility (IC) properties in the designed mechanisms~\cite{Ismail_FNWF_2022, ISMAIL_JSAC_2023_semantic}.

However, despite the great benefits of using semantic communication, enabling the rendering of the physical environment into the Metaverse platform is still facing serious challenges.
Specifically, ensuring the integrity of the transmitted data to the VSPs faces an emergent challenge due to the moral hazard problem, i.e., the risk of being dependent on the moral behavior of the other parties. Existing studies, e.g., \cite{ISMAIL_JSAC_2023_semantic, Minrui_EPViSA_2023}, are based on the general assumptions that data owners have no incentive to behave untruthfully after agreement on the prices and data to transmit. 
For instance, the authors in~\cite{Liew_TVT_2023} considered a blind trust on the content of the delivered semantic triplets by all SIoT devices. Nevertheless, this assumption is not always guaranteed.
A malicious adjustment by an attacker to the elements of the semantic triplet may cause the rendered digital twin to deviate from its physical world twin.
While incentive mechanisms with properties like IR and IC help prevent adverse selection issues before contract agreement, they do not fully address the moral hazard problem arising after agreement. Specifically, SIoT devices might deliver content deviating from the initial agreement with the VSP. For example, due to the cost incurred for data collection and extracting the semantics from these data at every timestep, a malicious sensing IoT device can choose to resend an old instance of the physical environment state repeatedly. Additionally, the rise of adversarial attacks based on AI generated content (AIGC)\footnote{https://petarpopovski-51271.medium.com/communication-engineering-in-the-era-of-generative-ai-703f44211933} complicates the distinction between genuine and manipulated content~\cite{Zhang_CCS_2023}. Particularly in the vehicular Metaverse, a malicious SIoT device could launch a poisoning attack on the delivered semantic data, disrupting the Metaverse system, leading to significant degradation of service quality (QoS) and user experience (QoE). Furthermore, there is a risk of life-threatening events that could trigger accidents in the physical world~\cite{Minrui_COMST_2023}.


From a theoretical perspective, the adverse selection problem can be mitigated if we can prove that there is an optimal solution for all the participants in the contract (or more generally, in the game) where the selection of any other action would negatively impact the misbehaving participant. However, that is not the case with the moral hazard issue. In moral hazard problems, the misbehaving party (e.g., SIoT device) has the objective to corrupt the system rather than increasing his monetary benefit. More specifically, the attack does not affect the selection of the appropriate bundle from the contract designer. Instead, the attack is done on the delivered content after the agreement on the contract between the two parties. Therefore, due to the versatility of moral hazard attacks, it is difficult to design a mechanism with theoretical proofs to prevent this serious type of attacks. Nevertheless, to ensure safe deployment of vehicular Metaverse, addressing the issue of moral hazard in these systems is critical and urgent. 

\subsection{Related Work}

\subsubsection{Poisoning Attacks}
Existing work address different types of poisoning attacks such as poisoning federated learning (FL) models where the objective of the attacker is to diverge the learning model towards unwanted or damaging outcomes.
The authors in~\cite{Tabatabai_2022_IWCMC}, present an evolutionary-based method for poisoning attack detection in federated learning models. The authors suggested clustering technique assembles clients into numerous clusters, each assigned a model chosen at random to capitalize on each model's unique strengths. Subsequently, the clusters are formed in a repetitive process, with the least effective cluster being removed after each iteration until only one cluster remains. Throughout these iterations, specific clients are excluded from clusters due to either using corrupted data or demonstrating subpar overall performance. The clients that demonstrate superior performance continue to participate in the subsequent iteration. The cluster that endures with these successful clients is then employed for training the most optimal FL model.

Poisoning attacks were also addressed in smart grid networks. For example, in~\cite{Alomrani_2023_TITS}, the authors addresses emerging cyber threats in electric vehicle charging networks where they highlighted how malicious actors can manipulate data to gain unauthorized charging privileges, disrupt schedules, and even pilfer power. To counter this, the paper introduces a detection model based on machine learning. The model is trained on both real-world driving data and a specially generated malicious dataset, crafted using a Reinforcement Learning (RL) agent. This agent is trained to devise sophisticated attacks that can evade basic detection methods while securing high charging priority for the malevolent EV.


\subsubsection{Reputation Mechanisms}
Reputation mechanisms are emerging as a highly efficient tools for quantifying trust among various entities in electronic markets, facilitating the establishment of trust and expediting cooperation.
For example, eBay's feedback mechanism is the key for enabling trustworthiness between geographically dispersed users over the Internet~\cite{Resnick_2002}.
Previous works used reputation as a tool to valuate the reliability of the participants in contracts~\cite{Kang_IoTJ_2019_Reputation, Yining_2011, Chrysanthos_2005_JSTOR}.
In~\cite{Chrysanthos_2005_JSTOR}, reputation mechanism was introduced with specific application on online trading from an economic perspective. The author studied several theoretical perspectives of different scenarios in online tradings.
In~\cite{Kang_IoTJ_2019_Reputation}, the authors combined contract theory with reputation mechanism to enable the FL server reliably select workers based on their historical interactions. However, the poisoning attack detection requires benign dataset to which the FL sever can compare with to discover the attacks. In our studied scenario, as new data is collected in real-time, there is no way to use this method to detect new poisoning attacks. Similar approaches where also adopted in~\cite{Uprety_2021_SSCI, Zunming_2022_mdpi} for FL. 

In the context of blockchain-based vehicular systems, the authors in~\cite{Zunming_2022_mdpi} used reputation scores to detect stragglers with low computing power and/or bad channel conditions from the FL training. The work in~\cite{YangZhe_2017} used a similar technique to detect credible users in vehicular network.
In~\cite{Piccolo_2023}, a decentralized reputation system was developed using consortium blockchain and smart contracts to verify the veracity of vehicular network alerts and detect malicious behavior in vehicular ad hoc networks (VANETs). By analyzing vehicle data reliability and implementing a comprehensive reputation mechanism, the proposed solution aimed to enhance driving safety by mitigating false message attacks.
In~\cite{Yining_2011}, the authors used subjective logic model to build trust amongst nodes in mobile ad-hoc networks (MANETs). As some nodes in the MANETs might have no willingness to fully cooperate, non-malicious nodes use a reputation mechanism to incentivize all other nodes to cooperate fully or face the risk of being expelled from the network.









\subsection{Contributions}
In this work, we address a new class of poisoning attacks with specific application in vehicular Metaverse. Specifically, the real-time collection of sensing data pose a challenging problem for poisoning attack detection as there is no baseline dataset to which the VSP can compare the collected data as in FL.
Importantly, existing studies on poisoning attacks assumes that the attacker has a unique behavior over their full interaction with the system. This assumption reduces the problem to detecting the attack and then prevent the attacker from any future interaction with the system.
However, malicious devices can have mixed behaviors over all their period of availability and these behaviors can be rooted to different causes. 
Additionally, in many applications, e.g., data collection for the Metaverse, we might be in need for all nearby SIoT devices to share their real-time data. Therefore, if we are able to correctly label the SIoT devices over all their period of interactions with the system, we can further maximize the benefice from all nearby SIoT devices.

To this end, we propose a heuristic algorithm based on subjective logic and reputation mechanism to reduce the risks of participants' misbehavior with specific application to vehicular Metaverse ecosystems. 
The centralization of indirect reputation enables a VSP to have deeper knowledge about the behavior of SIoT devices with other VSPs. This architecture indirectly disincentivize malicious SIoT devices from launching any type of attack on the delivered content.
In such contexts (i.e., \textit{moral hazard}), the goal of reputation mechanisms is to encourage cooperative and honest conduct among self-interested participants.
As far as our knowledge extends, prior research has not tackled this specific type of attacks on wireless communication ecosystems. 
We argue this by the fact that Metaverse services are bringing new and unprecedented perspectives and applications that has their specific set of problems that needs to be addressed rapidly.
To bridge this gap, our contributions aim to incentivize SIoT devices to consistently engage truthfully in Metaverse services and hence, ensure their safety against several types of attacks.

Part of this work was presented in~\cite{ismail_2024_WCNC} where we studied a scenario with single VSP interacting with several SIoT devices and vehicular Metaverse users (VMUs). The derived solution there is limited to introducing the backpropagation filtering algorithm to mitigate the poisoning attack.
In this paper, we extend our study to a more general scenario where we study the benefits of using a centralized reputation server to share reputations amongst several VSPs.
Additionally, unlike the previous work, which focused on filtering malicious inputs, this study ensures proactive deterrence through economic and reputational incentives.
Furthermore, the previous work lacked a structured framework for addressing poisoning attacks. In this extension, we introduce vehicular Metaverse guard (VMGuard), a comprehensive four-layer architecture, which incorporates reputation scoring via subjective logic, a vanishing mechanism to address false positives, and a real-time feedback loops between VMUs and VSPs to dynamically adjust reputations.
In summary, our primary contributions in this work are as follows:

\begin{itemize}
    \item We propose a new approach, namely VMGuard, that tackles the moral hazard issue in the vehicular Metaverse, offering a distinctive viewpoint on trust and security in this evolving technology. Then, we propose a reputation-based mechanism that acts as a deterrent against participant dishonesty, thereby bolstering the system's overall integrity.

    \item The proposed VMGuard framework is composed of four stacked layers that cooperate together to address several layers of the security of the vehicular Metaverse. The four layers are: reputation layer, semantic data collection layer, digital twin rendering layer and reputation backpropagation layer.
    
    \item By using subjective logic, we model the reputation of each SIoT device and implement a vanishing-like system to handle previous ratings. This innovative strategy aims to retain reputable SIoT devices within the market despite occasional false positive labels, preventing their premature exit.

    \item We study several strategies for managing the reputation memory and show the trade-offs for each strategy. We show that a mixed strategy that uses the most recent reputations in combination with a fading function yields the best performance across several anti-poisoning metrics.
\end{itemize}

The structure of the paper is as follows. In Section~\ref{sec:system_model} we presented our studied system model then in Section~\ref{sec:defense_model} we introduce our defense model.
We provide numerical results and insightful discussions about our VMGuard framework in Section~\ref{section_Results}. Section~\ref{section_conclusion} concludes the paper.

\section{System Model}\label{sec:system_model}
This section details the adopted system model and the different poisoning attack models and strategies.

\subsection{System Model Description}

\begin{figure}[ht!]
    \centering
    \includegraphics[width=.45\textwidth,height=4.5cm]{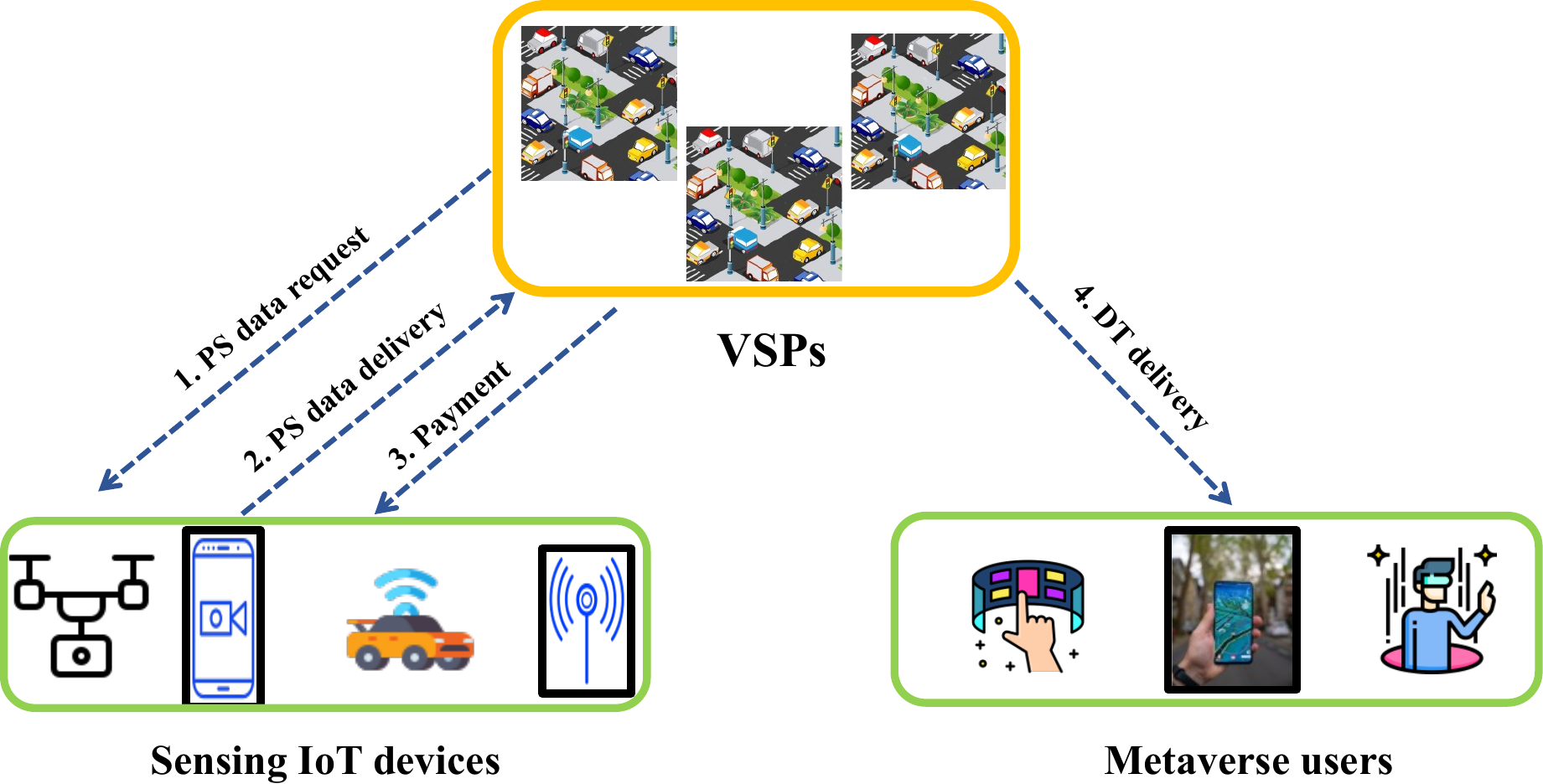}
    \caption{Studied system for poisoning attacks in vehicular Metaverse ecosystem.}
    \label{fig:studied_system}
\end{figure}

Fig.~\ref{fig:studied_system} describes our proposed system model. 
We examine a Metaverse market comprising a set of $\mathcal{L} = \{1, \dots, j, \dots, L\}$ VSPs, a set of $\mathcal{N} = \{1, \dots, i, \dots, N\}$ SIoT devices, acting as data owners and a set of $\mathcal{M} = \{1, \dots, m, \dots, M\}$ VMUs.
A VSP's primary role involves constructing the Metaverse's digital twin using data obtained from selected SIoT devices and delivering the constructed digital twin to the VMUs. Each SIoT device is equipped with sensors for geo-spatial data collection and machine learning models for extracting semantic information from raw data. These SIoT devices transmit extracted semantic data instead of the raw data, which is large in size, improving network efficiency~\cite{Ismail_FNWF_2022}. 
Additionally, the use of semantic communication has another major advantage on the system security. Specifically, this strategy eliminates any attempt to start an adversarial attack on the delivered images/videos. Although several work have been proposed to minimize the threat of adversarial attacks~\cite{xie2017mitigating, Zhang_CCS_2023}, adversarial attacks are still difficult to eliminate and having a system where such attacks cannot be implemented is advantageous. The attacks becomes restricted to poisoning attack, for which we show later our developed reputation-based mechanism to reduce its effect.

In the studied system, the VSPs start first by initiating requests to available SIoT devices to sell their semantic data for specific regions of the physical environment. 
SIoT devices can sell their semantic data to multiple VSPs simultaneously and therefore a multi-seller single buyer reverse auction mechanism can be adopted into our system design~\cite{Nisan_2007}. 
SIoT devices within the vicinity of each VSP express their interest in selling data by providing their offered prices. Each SIoT device $i$ conveys its type $t_i =\{b_i, \hat{s}_i\}$ to the VSP, where $b_i$ denotes the price for its data, and $\hat{s}_i$ represents the semantic value of the offered data. This semantic value, $\hat{s}_i$, as detailed in~\cite{Ismail_FNWF_2022}, incorporates factors such as weather conditions, object detection counts, and their respective positions. For instance, during adverse weather, the derived semantic information holds greater value, indicating its increased importance to the VSP.
Moreover, considering that SIoT devices need a particular number of channels to transmit their semantic data, each SIoT device $i$ informs the VSP about $C_i$, representing the required number of channels necessary for transmission.

Once the VSPs receive the offers from all SIoT devices, a reverse auction model is executed to select the set of winners (described later in Section~\ref{sec:defense_model}) by each VSP. Upon agreement between the VSPs and the SIoT devices, the SIoT devices start delivering their semantic data to the VSPs then receive back their payments. The VSPs then render their digital twins of the vehicular Metaverses based on the data received from the SIoT devices and deliver the DTs to the VMUs.
Finally, we consider that at every round of the auction, the number of participating SIoT devices and their bid values change due to SIoT devices' mobility or ability/inability to provide the requested semantic data.
As a defense model, several layers and protocols are implemented to mitigate poisoning attacks on vehicular Metaverse systems (detailed later in Section~\ref{sec:defense_model}).



\subsection{Poisoning Attack Models}


As the auction mechanism remains vulnerable to moral hazard participants performing poisoning attacks into the Metaverse system, real-time prevention of such attacks is crucial to maintain service integrity and high user experience.

\subsubsection{Poisoning Attack Description}
Various forms of poisoning attacks are possible from malicious SIoT devices. We outline here three primary types of these attacks:

\begin{itemize}
    \item \textbf{Reducing service cost:} As the continuous collection of sensing data and the execution of semantic data extraction algorithms require continuous energy consumption, a selfish SIoT device can choose to deliver semantic data that is based on outdated sensing data. This action will cause the malicious SIoT device to increase its utility (as lower energy consumption is required, see~\eqref{eq:u_i}) while the rendered digital twin will not reflect real-time dynamics of the physical world, causing degraded QoE of the VMUs.

    \item \textbf{Digital twin tampering:} While the previous type of attack does not have an objective of causing damages to the Metaverse services, the digital twin tampering is initiated by a malicious SIoT device to cause severe and targeted degradation of the Metaverse services. 
    For instance, a malicious device can hide an existing object in the physical world and only deliver an absence of obstacle which can cause a vehicle to clash with other objects causing serious security threats if the physical world is affected by actions taken in the digital world. Even if only the digital twin is affected, that would degrade the QoS and QoE of the Metaverse users.
    Fig.~\ref{fig:conf_semantci_attack} illustrates an instance of such attack.

    \begin{figure}[ht!]
        \centering
        \includegraphics[width=.4\textwidth,height=4.0cm]{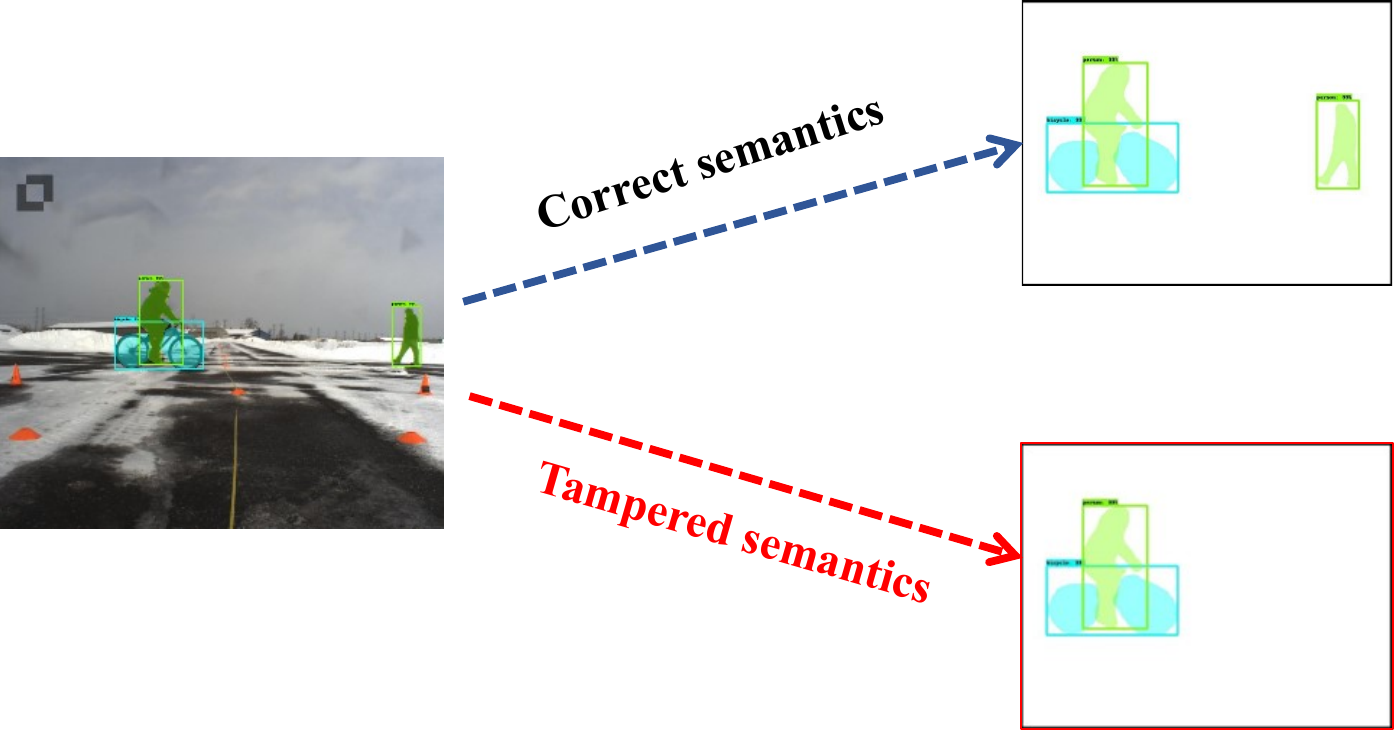}
        \caption{An example of a semantic attack where the attacker hides the person on the right side of the image~\cite{ismail_2024_WCNC}.}
        \label{fig:conf_semantci_attack}
    \end{figure}

    \item \textbf{Man-in-the-middle (MINM) attack:} This attack can be launched from an SIoT device with low security defense levels, i.e., not using higher level security schemes such as cryptography (already in the standard of 5G and most likely to remain in 6G \cite{Saad_6G_2020}). The attacker transmits their altered data instead of the legitimate SIoT device data. The VSP has no visibility of the existence of the attacker in the middle of the transmission. Therefore, if the attacker is detected, the compromised SIoT device is flagged as an untrustworthy device. Such a decision implies that the victim SIoT device will not be able to deliver its sensing data to the VSP which constitute a significant loss of valuable physical world information.
\end{itemize}

\subsubsection{Poisoning Attack Strategies}
At every time period, the attacker chooses either to attack by transmitting falsified data or not attacking by sending legitimate data. An attacker $i$'s objective is to maximize their accumulated utility defined as
\begin{equation}\label{eq:u_attacker}
    \sum^{T}_{t=1} u_i^t = \sum^{T}_{t=1} (p_i^t - c_i^t + d_i^t),
\end{equation}

\noindent where $T$ is the total number of time periods for the interaction between the SIoT devices and the VSP, $p_i^t$ and $c_i^t$ are the payment received from the VSP for their semantic data at time step $t$ and the cost for collecting and extracting the semantics at time step $t$, respectively. $d_i^t \in\{0, 1\}$ is a numerical that reflects the degree of damage caused by the attacker to the VSP's services at time step $t$. If the attack is successful $d_i^t=1$ and $0$ otherwise.

However, it is not trivial to calculate the parameters of \eqref{eq:u_attacker} and hence, hindering the use of any intelligent algorithm (e.g., deep reinforcement learning (DRL)) to maximize the attacker's profit.
Specifically, as the SIoT device has a limited observation about the environment, they cannot infer the exact reason for an SIoT device not being selected amongst the winners in the auction rounds following their attack. For example, it can be due to detection of the attack or simply because the capacity of the VSP to hire more SIoT devices was reached or newly arrived SIoT devices submitted higher bids than the attacker's bid. All these information are hidden from the attacker and deploying an intelligent algorithm, e.g., DRL-based model, would simply results in a random attack strategy.

Nevertheless, the attacker can still implement an opportunistic algorithm to maximize their revenue based on the partial information that they can infer. Specifically, if the attacker produce an attack and still get accepted in the next rounds of the semantic data requests, the attacker can infer then that their previous attack was successful and hence, infer that $d_i^t=1$. This encourages the attacker to continue their attack and increase their damage to the VSP's services. For clarity, we consider that malicious SIoT devices keep track of their previous successful attacks in a counter $d_i = \sum_{t=1}^{T} d_i^t$ and reformulate the utility function for all SIoT devices (i.e., both malicious and non-malicious SIoT devices) as follows
\begin{equation}\label{eq:u_i}
    u_i = p_i - c_i,
\end{equation}
\noindent where $p_i$ is the payment received from the VSP, and $c_i$ is the service cost which covers the communication cost, data collection cost and execution of the semantic extraction algorithms.

To sum-up, the general strategy of the attacker is to continuously attack until banned from providing semantic data.
The attacker then moves to the non-attack mode until accepted in future rounds of the auction. Once accepted again, the attacker can choose to follow an attack mode or a non-attack mode based on a predefined probability value $p_{atk}$. Later in Section~\ref{section_Results}, we show how changing the value of $p_{atk}$ affects the strategy of the attacker and their revenue.



\section{Defense Model}\label{sec:defense_model}
    
\begin{figure}[ht!]
    \centering
    \includegraphics[width=.45\textwidth,]{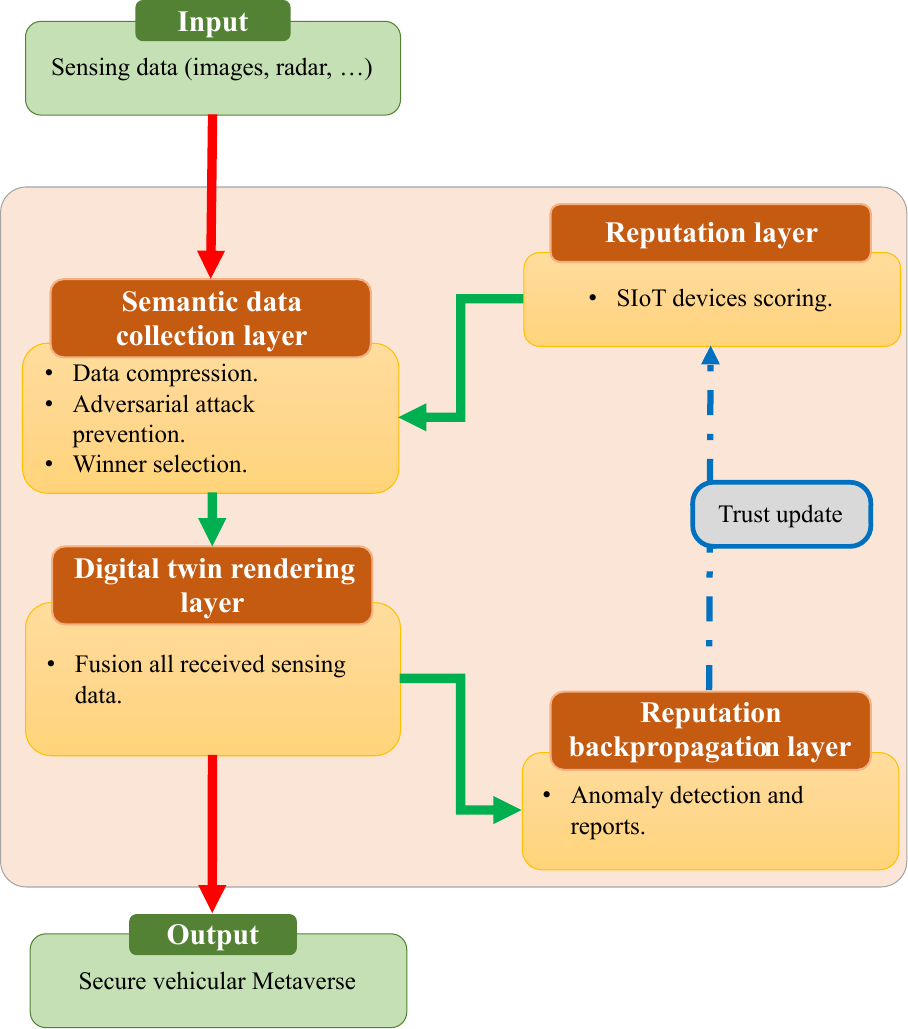}
    \caption{General workflow of VMGuard.}
    \label{fig:general_workflow}
\end{figure}

\begin{figure*}
\centering
     \begin{subfigure}[b]{0.49\textwidth}
         \centering
         \includegraphics[width=\textwidth]{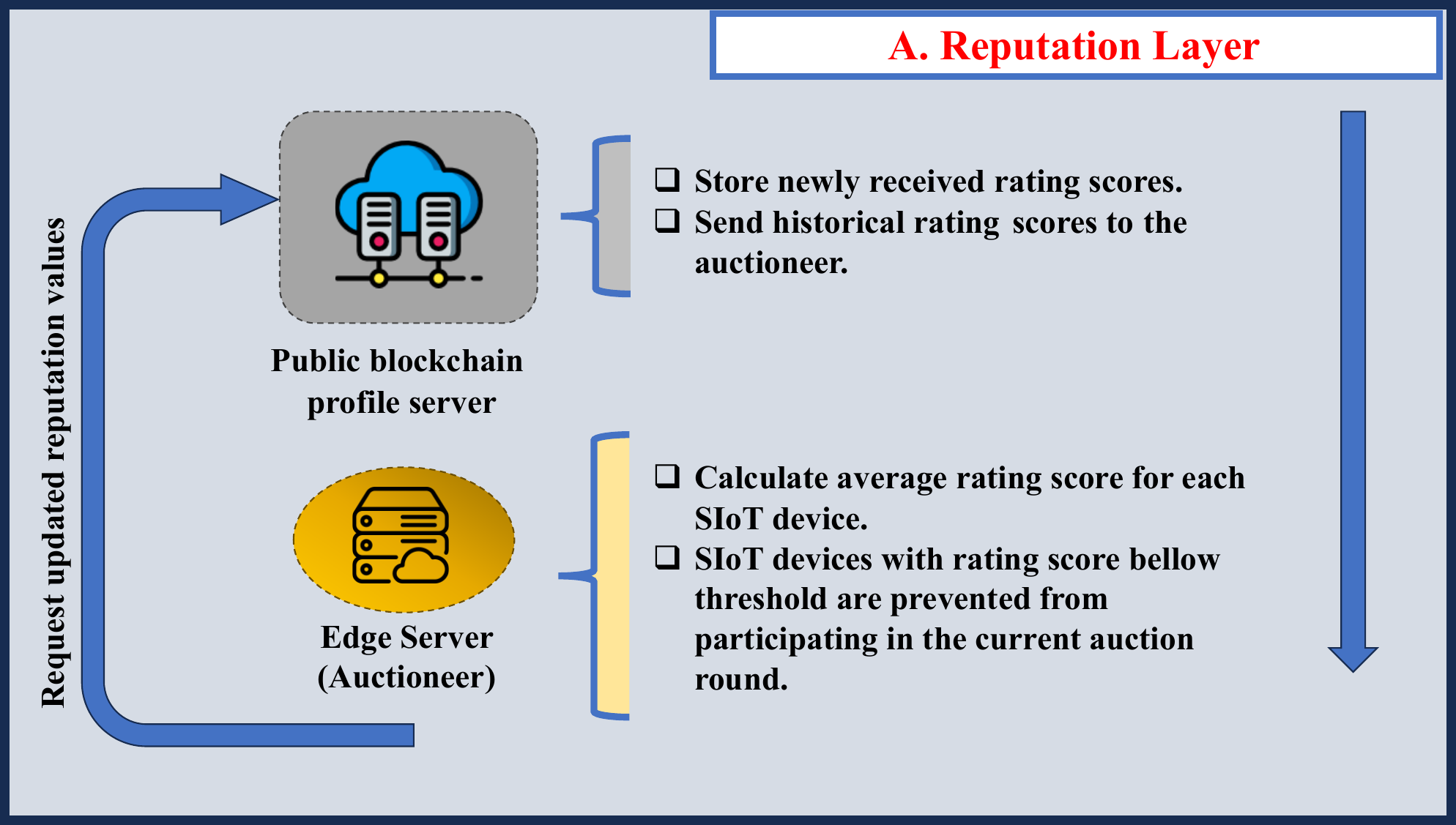}
         \caption{}
         \label{fig:A_Layer}
     \end{subfigure}
     \hfill
     \begin{subfigure}[b]{0.49\textwidth}
         \centering
         \includegraphics[width=\textwidth]{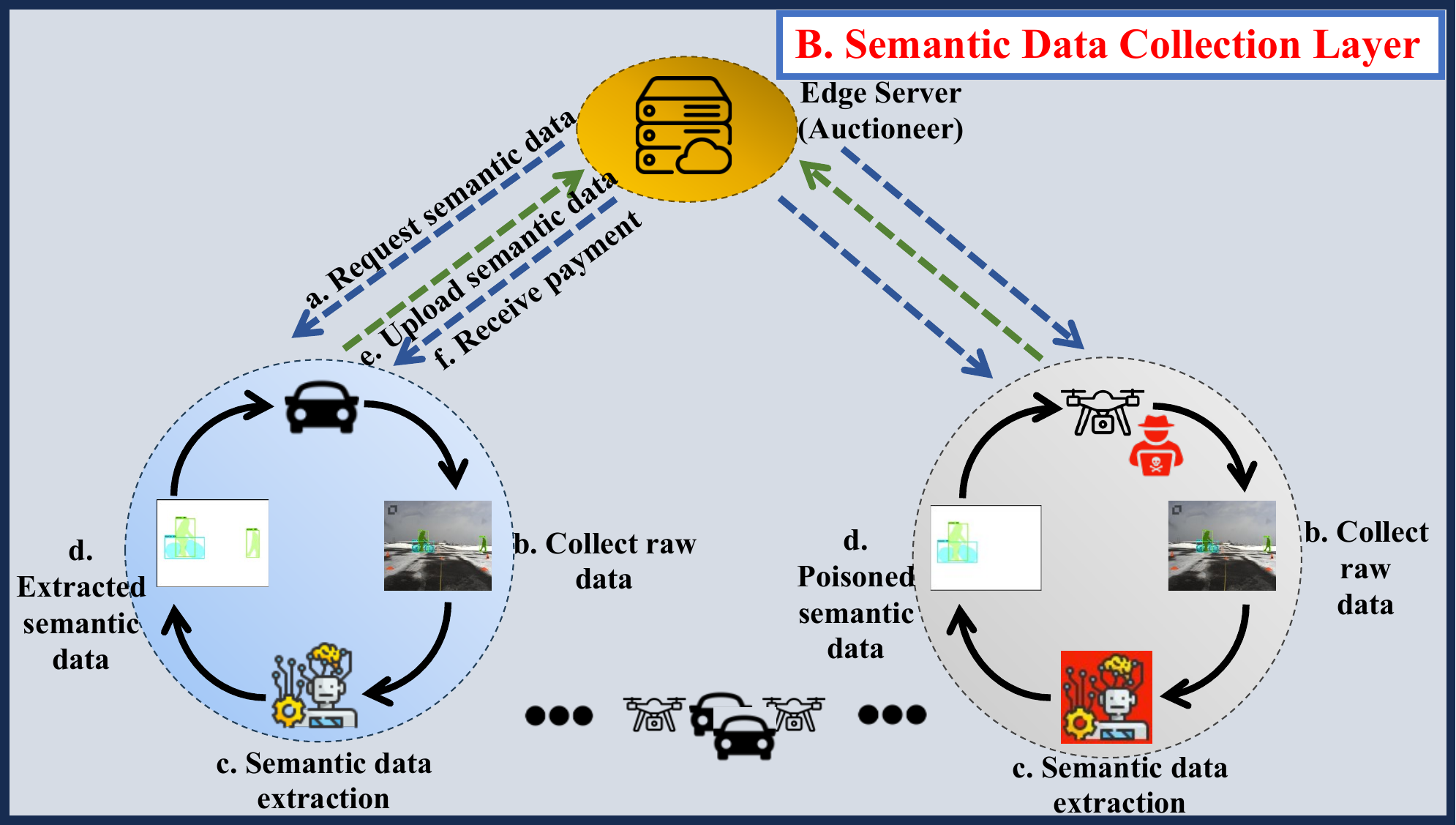}
         \caption{}
         \label{fig:B_Layer}
     \end{subfigure}
     \hfill
     \begin{subfigure}[b]{0.49\textwidth}
         \centering
         \includegraphics[width=\textwidth]{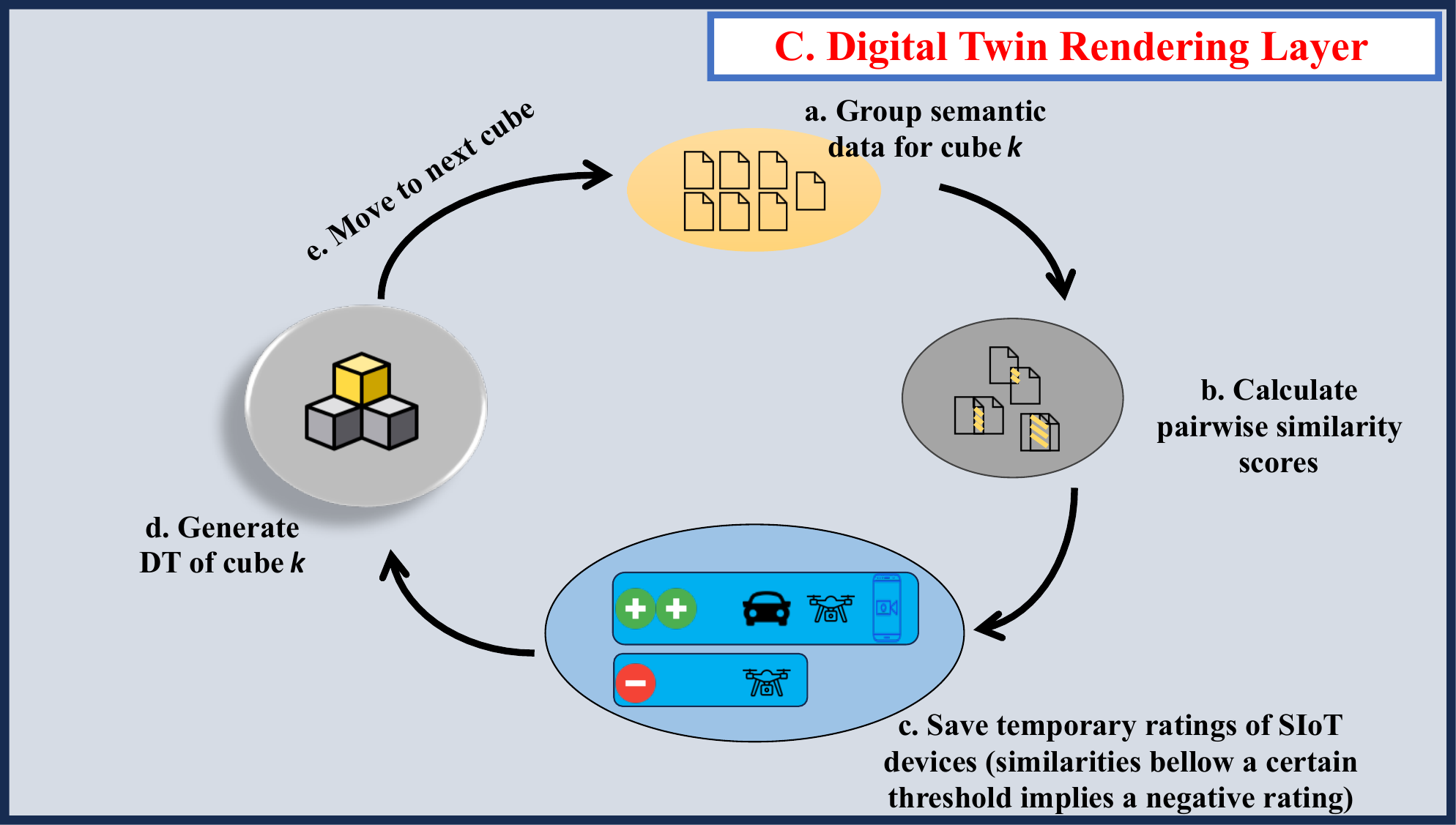}
         \caption{}
         \label{fig:C_Layer}
     \end{subfigure}
     \hfill
     \begin{subfigure}[b]{0.49\textwidth}
         \centering
         \includegraphics[width=\textwidth]{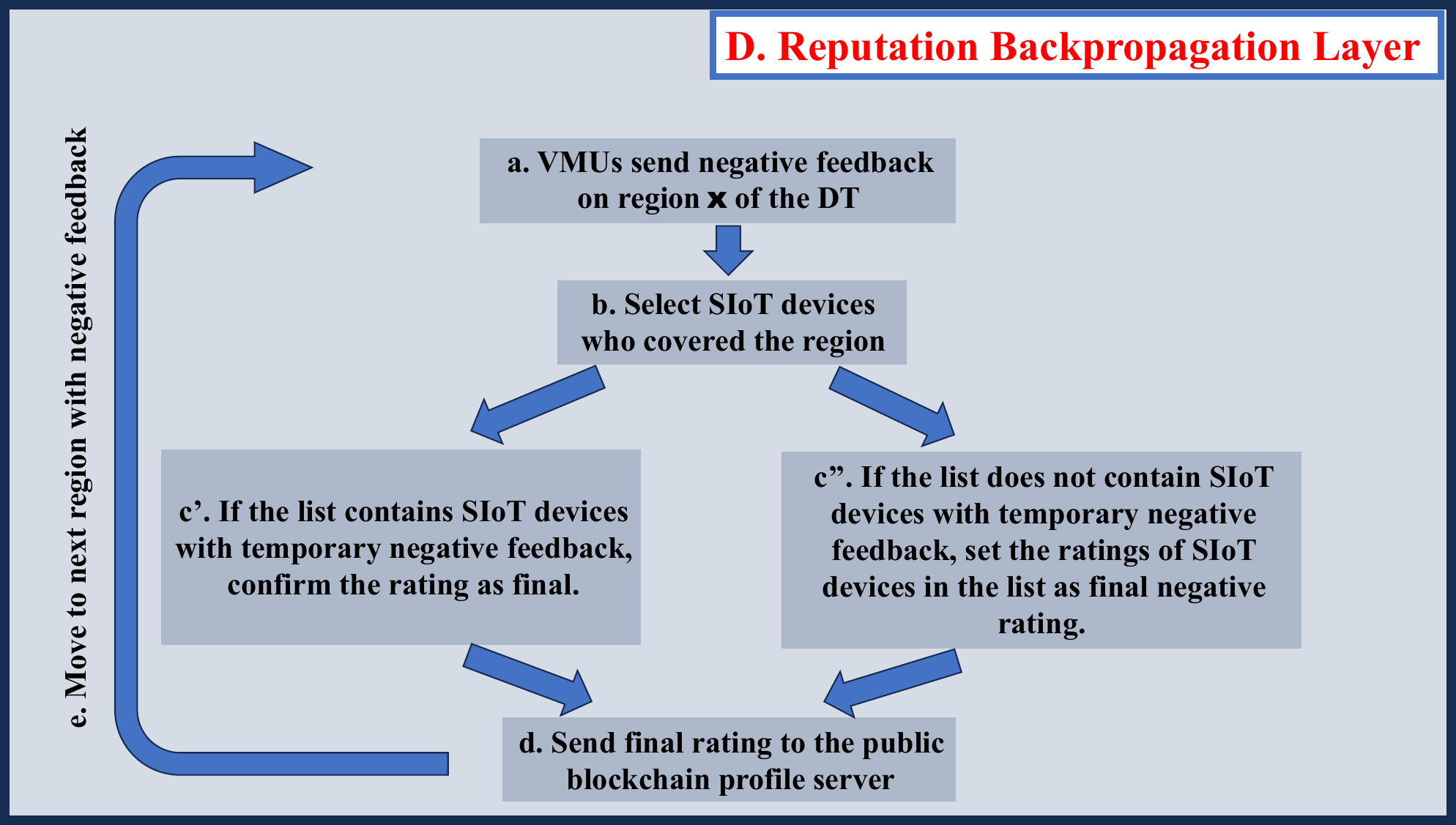}
         \caption{}
         \label{fig:D_Layer}
     \end{subfigure}
        \caption{Detailed description of VMGuard framework.}
        \label{fig:General_Figure}
\end{figure*}

As discussed in Section~\ref{sec:system_model}, the studied system model illustrated in Fig.~\ref{fig:studied_system} is vulnerable to poisoning attacks by malicious SIoT devices. To mitigate this serious security issue, we propose the defense model, namely vehicular Metaverse guard (VMGuard), which is presented in Fig.~\ref{fig:general_workflow} and detailed in Fig.~\ref{fig:General_Figure}.
VMGuard enables real-time protection against poisoning attacks through a four-layer architecture: reputation layer, semantic data collection layer, digital twin rendering layer and reputation backpropagation layer. The semantic data collection layer takes raw data as input and communicate with the reputation layer to check if the SIoT device is trustworthy or not. Once the semantic data is ready, the SIoT device send the semantic data to the VSP where the digital twin rendering layer is implemented to create the digital copy of the physical world by merging all the received data from the SIoT devices. At this stage, the reputation backpropagation layer checks if there is any tempered data sent by the SIoT devices. If such tampering is detected, the system backpropagates the trust scores to the reputation server.
This integrated approach provide a robust and real-time detection system that can effectively prevent poisoning attacks on vehicular Metaverse systems.

In what follows, we describe the four layers in detail and show how they do interact with each others to enhance the system security.

\subsection{Reputation Layer}


Different from the naive system model presented in Fig.~\ref{fig:studied_system}, we consider that the communication between the SIoT devices and the VSPs is enabled through an edge server which also plays the role of the auctioneer (see Fig.~\ref{fig:General_Figure}). Additionally, we consider a public blockchain profile server responsible for storing the new ratings for each SIoT device which are based on the quality of the delivered semantic data to the VSPs.
The use of blockchain in the public profile server is motivated by the structure of blockcahin technology which is a decentralized and tamper-proof distributed shared ledgers and databases. Blockcahins operate openly, offering anonymity and traceability features~\cite{Yutao_TPDS_2019, Kang_IoTJ_2019_Reputation}.

Reputation layer is the first layer in our VMGuard framework. In the reputation layer, historical reputations of all SIoT devices are retrieved from the public blockchain profile server by the edge server. Both the public blockchain profile server and the edge server play key roles in the design of our framework. 
The edge server is considered as a hub connecting all the SIoT devices to all the VSPs. The edge server constantly broadcasts requests for semantic data from the VSPs to all the SIoT devices.
The use of a centralized public reputation server has a crucial role in enhancing the security of the vehicular Metaverse system. Specifically, this enables exploiting benefits from VMUs interaction with different SIoT devices and hence, early prevention of  damages from potential malicious SIoT devices. This strategy is more efficient compared to the case where every VSP keeps their interactions with SIoT devices private~\cite{Kang_IoTJ_2019_Reputation}.

Next, interested SIoT devices send their offers to provide each VSP with the requested sensing data (e.g., images, GPS, Lidar and radar signals) in a form of extracted semantic data.
After receiving the offers from the SIoT devices, the edge server checks their reputations using a subjective logic model (described bellow) based on data fetched from the public blockchain profile server. The subjective logic model outputs a reputation value which needs to be above a certain threshold $\delta$ to consider the SIoT device to have a positive rating. Otherwise, the rating is marked as negative and the flagged SIoT device will not be allowed to participate in the auction process. SIoT devices with positive rating are stored in sets $\mathcal{N}_j$ for each VSP $j$ and used in the winner selection algorithm to decide the final list of SIoT devices from which semantic data is collected. The edge server acts also as the auctioneer where the winner selection algorithm is executed to determine the winners. Once the set of winners is derived, the edge server notifies selected SIoT devices (winners) to start transmitting their semantic data.

Note that the reputation values are stored in a an open blockchain ledger to further discourage malicious devices from launching poisoning attacks but from different perspectives. Specifically, in case all the VSPs solicit an edge server to perform the tasks of the reputation layer (which is the same as our proposition as in Fig.~\ref{fig:A_Layer}), a malicious SIoT device would recognize that misbehavior with one VSP would result in real-time reputation penalties from other VSPs in subsequent interactions.
If the VSPs opt not to use a shared edge server as the auctioneer, each VSP must gather reputation values from the public blockchain profile server and compute average reputations for each SIoT device. Consequently, once an SIoT device is labeled as untrustworthy by one VSP, this information propagates through the public blockchain profile server, dissuading the flagged device from further misbehavior with other VSPs.

\subsubsection{Subjective Logic Model for Reputation Calculation}

The public blockchain profile server sends historical rating scores to the edge server (acting also as the auctioneer) to calculate new reputation scores for every SIoT device.
Based on the framework of subjective logic~\cite{Yining_subjective_2009}, we use the tuple $r_{i:j}=\{b_{i:j}, d_{i:j}, u_{i:j}, \alpha_{i}\}$ to denote elements of the reputation score calculated based on feedbacks from VSP $j$ about SIoT device $i$.
$b_{i:j}$, $d_{i:j}$, $u_{i:j}$ and $\alpha_{i}$ represent belief, disbelief, uncertainty, and the effective degree of uncertainty effect on reputation of SIoT device $i$, respectively~\cite{Kang_IoTJ_2019_Reputation, Oren_subjctv_2007}. 
$b_{i:j}+ d_{i:j}+ u_{i:j} = 1$ and $b_{i:j}$, $d_{i:j}$, $u_{i:j}$, $\alpha_{i} \in [0,1]$.
We define $R_{i:j}$ as the reputation of SIoT device $i$ based on the tuple derived from VSP $j$'s feedback and is given by
\begin{equation}
    R_{i:j} = b_{i:j} +\alpha_{i}u_{i:j}.
\end{equation}

Next, we show how the elements of the tuple $r_{i:j}$ are calculated by the edge server.
As the VSPs interact with SIoT devices and receive feedbacks from VMUs over several time periods, the reputation for an SIoT device $i$ constructed at the edge server at each time period is expressed for this period. 
The effective period of interaction between the VSPs and the SIoT devices $T$ is divide into $Y$ time windows, i.e., $\{t_1, \dots, t_y, \dots, t_Y\}$. The reputation value of SIoT device $i$ based on feedback from VSP $j$ at time window $t_y$ is then calculated as

{\small \begin{equation}\label{eq:reputation_naive}
\begin{cases} 
b^{t_y}_{i:j} = \frac{\omega_1 p^{t_y}_{i:j}}{\omega_1 p^{t_y}_{i:j} + \omega_2 q^{t_y}_{i:j}+\kappa},
\\ d^{t_y}_{i:j} = \frac{\omega_2 q^{t_y}_{i:j}}{\omega_1 p^{t_y}_{i:j} + \omega_2 q^{t_y}_{i:j}+\kappa},
\\ u^{t_y}_{i:j} = \frac{\kappa}{\omega_1 p^{t_y}_{i:j} + \omega_2 q^{t_y}_{i:j}+\kappa},
\end{cases}
\end{equation}}

\noindent where $p^{t_y}_{i:j}$ and $q^{t_y}_{i:j}$ represent the number of all positive and negative interactions between the SIoT device $i$ and VSP $j$ at time window $t_y$, respectively. $\omega_1$ and $\omega_2$ are weighting factors for positive and negative interactions, respectively, and are considered to sum-up to 1, i.e., $\omega_1+\omega_2=1$. $\kappa$ is a constant value that reflects the rate of uncertainty and is set to $\kappa=1$~\cite{Kang_IoTJ_2019_Reputation, Oren_subjctv_2007}.

\subsubsection{Reputation Memory Management}
However, the reputation calculation described in \eqref{eq:reputation_naive} does not differentiate between old and recent feedbacks which can significantly affect the decision about the trustworthiness of an SIoT device. Specifically, SIoT devices can change their behavior at any point of interaction. Consequently, the system should be well designed to reflect the importance of recent interaction compared to old interactions. 
Recent interaction events hold greater significance due to their freshness, carrying a higher weight than those from the past.
A straightforward reputation memory management as formulated in \eqref{eq:reputation_naive} would be less efficient as we might penalize SIoT devices forever and lose a considerable volume of legitimate semantic data.
Therefore, we propose the following strategies to model this metric.

\begin{itemize}
    \item \textbf{Fading Function Based Strategy (\textit{Strategy 1}):} Here, we define the fading function that reflects this behavior as $ \psi(t_y) = \psi_y = z^{Y-y}$, where $z\in(0,1)$ is a given fading parameter, and $y$ is the current time window that determines the freshness degree of the reputation value. Therefore, \eqref{eq:reputation_naive} is reformulated as

    {\small \begin{equation}\label{eq:reputation_fading}
    \begin{cases} 
    b^{fin}_{i:j} = \frac{\sum^{Y}_{y=1}\psi_y b^{t_y}_{i:j}}{\sum^{Y}_{y=1}\psi_y},
    \\ d^{fin}_{i:j} = \frac{\sum^{Y}_{y=1}\psi_y d^{t_y}_{i:j}}{\sum^{Y}_{y=1}\psi_y},
    \\ u^{fin}_{i:j} = \frac{\sum^{Y}_{y=1}\psi_y u^{t_y}_{i:j}}{\sum^{Y}_{y=1}\psi_y}.
    \end{cases}
    \end{equation}}
    
    Therefore, the final reputation value of SIoT device $i$ from VSP $j$ is calculated as $R^{fin}_{i:j} = b^{fin}_{i:j} +\alpha_{i}u^{fin}_{i:j}$.
    Finally, we calculate the final reputation value of an SIoT device $i$ as the average value of the reputation values received from all VSPs, i.e., 
    \begin{equation}\label{eq:R_fin}
        R^{fin}_{i} = \frac{\sum^L_{j=1}R^{fin}_{i:j}}{L}.
    \end{equation}

    \item \textbf{Most Recent Feedback Strategy (\textit{Strategy 2}):} Here, we limit the observation of the rating values to the most recent $D$ ratings across all time windows. Therefore, \eqref{eq:reputation_naive} is reformulated as follows

    {\small \begin{equation}\label{eq:reputation_D_recent}
    \begin{cases} 
    b^{D}_{i:j} = \frac{\omega_1 p^{D}_{i:j}}{\omega_1 p^{D}_{i:j} + \omega_2 q^{D}_{i:j}+\kappa},
    \\ d^{D}_{i:j} = \frac{\omega_2 q^{D}_{i:j}}{\omega_1 p^{D}_{i:j} + \omega_2 q^{D}_{i:j}+\kappa},
    \\ u^{D}_{i:j} = \frac{\kappa}{\omega_1 p^{D}_{i:j} + \omega_2 q^{D}_{i:j}+\kappa},
    \end{cases}
    \end{equation}}

    \noindent where $p^{D}_{i:j}$ and $q^{D}_{i:j}$ represent the number of positive and negative interactions between the SIoT device $i$ and VSP $j$, respectively, for the most recent $D$ interactions.
    The final reputation value of SIoT device $i$ from VSP $j$ is calculated in this case as $R^{D}_{i:j} = b^{D}_{i:j} +\alpha_{i}u^{D}_{i:j}$ and the average value of the reputation values received from all VSPs is calculated as
    \begin{equation}\label{eq:R_fin_D}
        R^{D}_{i} = \frac{\sum^L_{j=1}R^{D}_{i:j}}{L}.
    \end{equation}

    \item \textbf{Mixed Strategy (\textit{Strategy 3}):} Here we use a fading function similar to that in \textit{Strategy 1} and limit the access to the reputation values in the public profile server to the most recent $D$ time windows as in \textit{Strategy 2}. Therefore, \eqref{eq:reputation_naive} is reformulated as follows

    {\small \begin{equation}\label{eq:reputation_mixed}
    \begin{cases} 
    b^{finD}_{i:j} = \frac{\sum^{Y}_{y=Y-D}\psi_y b^{t_y}_{i:j}}{\sum^{Y}_{y=Y-D}\psi_y},
    \\ d^{finD}_{i:j} = \frac{\sum^{Y}_{y=Y-D}\psi_y d^{t_y}_{i:j}}{\sum^{Y}_{y=Y-D}\psi_y},
    \\ u^{finD}_{i:j} = \frac{\sum^{Y}_{y=Y-D}\psi_y u^{t_y}_{i:j}}{\sum^{Y}_{y=Y-D}\psi_y}.
    \end{cases}
    \end{equation}}

\end{itemize}


Once the reputations of all SIoT devices are derived for the current time slot (based on one of the above described strategies), the framework moves to the next layer, i.e., semantic data collection layer.





\subsection{Semantic Data Collection Layer}

Upon reception of the list of SIoT devices with positive reputation from the reputation layer (i.e., reputation value above threshold $\delta$), the edge server initiates the semantic data collection process. The edge server (acting also as the auctioneer) starts first by executing the reverse auction algorithm to select the set of SIoT devices from which to buy the semantic data. Once the set of winners is derived for each VSP, the edge server requests the winning SIoT devices to deliver their semantic data (as described in Fig.~\ref{fig:B_Layer}, step a). 
The SIoT devices then proceed to collect raw data from the physical world (Fig.~\ref{fig:B_Layer}, step b) and then provide the collected data to the ML algorithm to extract the semantic data (Fig.~\ref{fig:B_Layer}, step c). 
At this stage, an attacker might decide to launch a poisoning attack on the VSP by falsifying the semantic data (Fig.~\ref{fig:B_Layer}, step d). For instance, the attacker can remove an object from the collected scene and lure the VSP about the existence of such object (pedestrian in step d, Fig.~\ref{fig:B_Layer}). 
The main objective of our VMGuard framework is to mitigate such attacks.
Finally, once the semantic data is received by the VSPs, payments are given to the SIoT devices based on the agreed prices.

Note here that an intuitive solution to the moral hazard problem would be to delay the payment until the VSP ensures that the received semantic data is not falsified. Nevertheless, although the VSP will not suffer from payment loss, the time and resources used to receive the semantic data from a malicious device are also lost as the VSP would be able to buy semantic data from other non-malicious sensing IoT devices. For this reason, delaying the payment is not an efficient solution to this moral hazard problem.

\subsubsection{Social Welfare Maximization Reverse Auction}



The social welfare, denoting the overall efficiency of the Metaverse market system concerning the sale of semantic data by SIoT devices to VSPs, is a pivotal metric for system evaluation~\cite{Zhang_2017_welfare}. Consequently, maximizing social welfare is indicative of the system's efficient operation. A comprehensive explanation of our developed reverse auction mechanism, addressing these considerations, is available in~\cite{Ismail_FNWF_2022}. In this context, we briefly outline the utility of SIoT devices and VSPs to formulate the social welfare of the auction mechanism for each VSP, a crucial aspect for performance evaluation.

The VSP's utility is defined as 
\begin{equation}\label{eq:u_hat}
    \hat{u} = \sum\limits_{i \in \mathcal{N}}\xi_i R_i^S  - \sum\limits_{i\in \mathcal{N}}\xi_i p_i - \hat{c},
\end{equation}

\noindent where $\xi_i$ is a binary variable denoting whether an SIoT device $i$ is selected amongst the winners or not, $R_i^S$ is the value of the semantic data delivered to the VSP by SIoT device $i$ and $\hat{c}$ is the cost for allocation of a wireless channel from spectrum service providers by the VSP. The system's social welfare is expressed as the sum of utilities generated by all entities within the system, encompassing both the VSP and the SIoT devices. This is mathematically written as
\begin{equation}
    S(\xi) = \sum\limits_{i\in \mathcal{N}}\xi_i \left( R_i^S - c_i \right) - \hat{c}.
\end{equation}    

Finally, the social welfare maximization problem can be written as an integer linear programming (ILP) subject to the availability of channels as follows
\begin{subequations}
\label{eq:optz_1}
\begin{align}
\begin{split}
\max_{\xi} S(\xi) = \sum\limits_{i\in \mathcal{N}}\xi_i \left( R_i^S - c_i \right) - \hat{c}, \label{eq:MaxA} 
\end{split}\\
\begin{split}
\hspace{1cm} s.t. \sum_{i\in \mathcal{N}} \xi_i C_i \leq B ,\label{eq:MaxB}
\end{split}\\
\begin{split}
\hspace{1cm} \xi_i \in \{0,1\}, \forall i\in \mathcal{N}, \label{eq:MaxC}
\end{split}
\end{align}
\end{subequations}

\noindent where $C_i$ is the number of channels requested by SIoT device $i$ and $B$ represents the number of channels offered by the VSP. The ILP presented in~\eqref{eq:optz_1} can be straightforwardly solved using a deterministic off-the-shelf solver, such as the \emph{Gurobi optimizer}.

\subsection{Digital Twin Rendering Layer}
Upon reception of the semantic data by the VSP, each VSP starts rendering the digital twin of their area of interest. Although the use of historical reputations is expected to prevent a number of attacker to deliver their falsified data, new attackers with no historical attacks or considerably old attacks can still bypass the reputation layer and deliver their data to the VSPs. To detect new attackers, we propose that the VSP implements a real-time detection mechanism during the digital twin rendering phase followed by a human-in-the-loop strategy denoted as the reputation backpropagation layer.

In~\cite{Shayan_2018}, the authors proposed the Reject on Negative Influence (RONI) technique to detect poisoning attacks on federated learning models. The concept involves assessing the impact of a local model on a predetermined database provided by the task publisher. This evaluation includes comparing the effects when using the local model against the case where the local model is not employed.
If the local model's performance drops below a predefined threshold set by the system when updating the database, this specific local model update will be declined upon integrating all local model updates.
However, what make it more challenging for us is that in our settings, the environment is fast changing making the use of a pre-defined database about the environment status not feasible.
Additionally, it is not trivial to check for signs of manipulation or tampering of the delivered content without having a baseline.

To address the aforementioned limitations, we propose the following strategy (illustrated in Fig.~\ref{fig:C_Layer}). 
In the 3D digital twin render layer, each VSP starts first by grouping semantic data for each cube of the 3D grid (Fig.~\ref{fig:C_Layer}, step a). Next, the VSP goes through all the semantic data for each cube and calculate a pairwise similarity score between semantic data received from all SIoT devices (Fig.~\ref{fig:C_Layer}, step b).
As the semantic data can be encoded in several ways and based on the type of which the semantic data is encoded, the similarity between two sets of semantic data can be calculated accordingly. For example, in~\cite{Ismail_FNWF_2022} the semantic data is encoded into JSON text files while in~\cite{Liew_TVT_2023} the semantic data is encoded into triplets of scene graphs, i.e., (“object 1”,“relation”,“object 2”)\footnote{The $L_p$ norms ($L_0$, $L_2$, $L_{\infty}$) represent the typical distance measures frequently used to measure similarity and can be used here. For instance, several off-the-shelf tools are available online to calculate the similarity between two JSON files, e.g., https://www.jsondiff.com/. }.

Based on the derived similarity scores, the VSP groups the SIoT devices in ascending order. SIoT devices which have similarity scores bellow a certain threshold $\gamma$ are attributed a temporary negative flags as they are considered potential attackers. The remaining SIoT devices with similarity scores above threshold $\gamma$ are attributed a temporary positive flags (Fig.~\ref{fig:C_Layer}, step c). Next, the algorithm generate the 3D digital twin of the current cube then moves to the next cube to process its data (Fig.~\ref{fig:C_Layer}, steps d and e). Once all the cubes in the 3D grid are examined, the VSP moves to the final step and generate the 3D environment and delivers the digital twin to the VMUs\footnote{Several studies exist in the literature for digital twin rendering based on text data and/or images, e.g., \cite{Gammeter_2010, yeh_2017_semantic, Zoltan_2018} . Also, as AIGC is emerging recently, the semantic data can be used an an input for large language models (LLMs) to generate 3D Metaverse world, e.g., as in~\cite{Hongyang_2023_TMC_Metaverse}.}.

\subsection{Reputation Backpropagation Layer}
Motivated by the paradigm of human-in-the-loop where humans' feedback is introduced as part of the system design, we propose the \emph{reputation backpropagation mechanism}.
Specifically, as the QoE of the VMUs is directly impacted by the quality of the digital twin, the VMUs can provide rich set of information to the VSPs from which we can infer the occurrence of poisoning attacks. For instance, if a VMU finds itself in a car accident situation in the Metaverse, while no information about existing obstacle in the road was provided, we directly infer that the provided digital twin is corrupted due to a poisoning attack. Therefore, we design  the reputation backpropagation layer to make use of the VMUs feedbacks to validate the temporary ratings provided as outputs from the digital twin rendering layer.

However, as there is no direct mapping between the VMUs and the SIoT devices, an important issue is how to backpropagate the feedback received by a VMU at the VSP server to a specific SIoT device. Motivated by the idea of backpropagation in deep neural networks (DNN)\cite{Goodfellow_2016}, we develop a new approach to backpropagate the feedbacks received by the VSPs from the VMUs. The reputation backpropagation is conducted as follows:

\begin{itemize}
    \item Upon reception of a negative feedbacks on some regions in the digital twin (Fig.~\ref{fig:D_Layer}, step a), each VSP brings out the list of SIoT devices that covered the area of interest (Fig.~\ref{fig:D_Layer}, step b).

    \item If the list contains SIoT devices with temporary negative feedbacks, we confirm the rating as final positive ratings (Fig.~\ref{fig:D_Layer}, step c').
    Otherwise, if the list does not contain SIoT devices with temporary negative feedback, we set the ratings of SIoT devices in the list as final negative ratings (Fig.~\ref{fig:D_Layer}, step c'').

    \item Finally, the VSP sends final ratings to the public reputation blockchain server (Fig.~\ref{fig:D_Layer}, step d).
\end{itemize}

The VSP then moves to the next region which received negative ratings until all the regions are covered (Fig.~\ref{fig:D_Layer}, step e). The newly derived ratings are used in the next rounds of the VMGuard framework (i.e., reputation layer)\footnote{Note here that while we do not directly measure QoE in the current work, it can be inferred from the system's ability to maintain low rates of tampered data and the successful integration of reliable SIoT devices. Additionally, future studies could introduce direct QoE metrics, such as user satisfaction ratings or response latency in real-time Metaverse services.}.

\section{Numerical Evaluation}\label{section_Results}
Within this section, we conduct experiments utilizing real-world data and present numerical outcomes to evaluate the efficacy of our suggested reputation-based incentive mechanism in countering poisoning attacks. Unless specified differently, we assume that there are $N=20$ SIoT devices, and each device has a fixed channel capacity of $r= 10 kbps$.

\subsection{Dataset and Simulation Settings}
Our experiments utilize the CARRADA dataset~\cite{Ouaknine_CARRADA_2020}, a contemporary open dataset featuring 30 scenes of synchronized sequences comprising camera and radar images. The radar images include both the range-angle (RA) and range-Doppler (RD) views.
For a comprehensive dataset description, we refer the reader to~\cite{Ismail_FNWF_2022}.
The outcome of the semantic segmentation algorithms produces images showcasing masked objects against a white background (refer to Fig.~\ref{fig:conf_semantci_attack}). Furthermore, a meta-data text file in JSON format is generated, mapping each object within the image to its corresponding class, along with additional derived semantic attributes such as range and shape.
The meta-data text file is limited to a maximum size of $d_{meta} = 1 kb$.
We use an online tool to calculate the similarities between the content of two JSON files\footnote{https://www.jsondiff.com/}.
To simulate real-world conditions where distinct SIoT devices capture identical scenes from varied angles, we assume that these devices transmit sets of scenes extremely close in time (measured in $ms$). Each time window involves utilizing a different scene from the synchronized sequences of 30 scenes\footnote{The implementation of the 3D grid representation remains a potential aspect for future development.}.
Moreover, our simulations span 10 time windows, each comprising 50 time-steps.

The attackers adopt a randomized strategy when targeting the vehicular Metaverse system~\cite{Sakib_2023}. In this scenario, a malicious SIoT device intermittently switches between transmitting genuine semantic data and transmitting fabricated semantic data, intending to deceive the VSP based on their probability of attack $p_{atk}$.
The attacks are implemented by randomly dropping some masked objects in addition to any related content in the JSON file associated with the image. 
To model the different types of SIoT devices that participate in the vehicular Metaverse market, we group the SIoT devices into three types:
\begin{itemize}
    \item \textbf{Type-0 SIoT:} This type reflects normal SIoT which always transmit legitimate semantic data, i.e., no willingness to attack and no man-in-the-middle attack.
    

    \item \textbf{Type-1 SIoT:} at every time window, an attack is launched with probability $p_{atk}=0.5$ in each timestep.

    \item \textbf{Type-2 SIoT:} at the beginning of every time window, choose with probability $p_{atk}=0.5$ either to attack for the whole time window or not to attack.
\end{itemize}

\subsection{Results}

\subsubsection{Performance of VMGuard Against Different SIoT devices Types}
\begin{figure}[ht!]
    \centering
    \includegraphics[width=.45\textwidth,height=4.5cm]{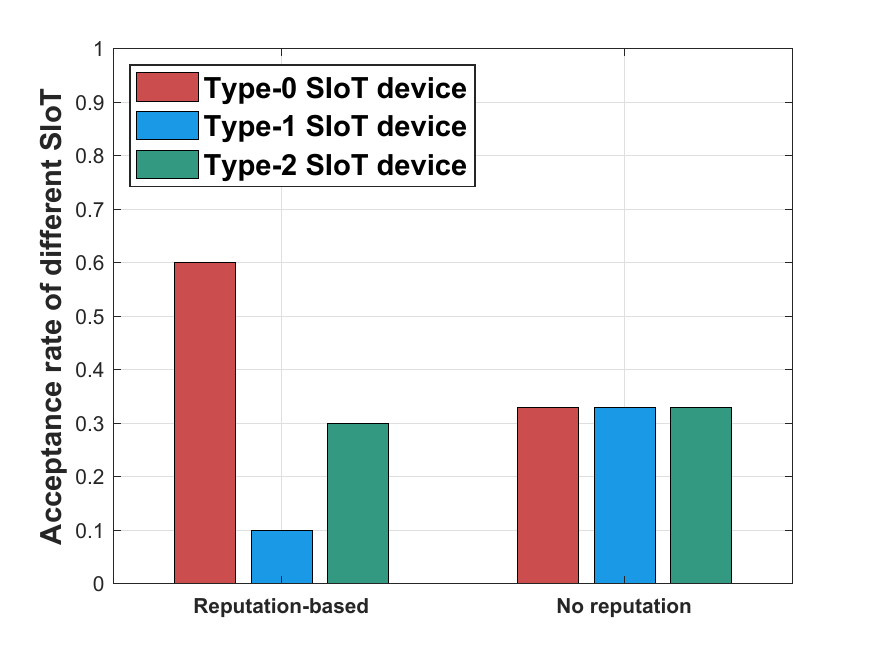}
    \caption{Acceptance rate of SIoT devices from different types with and without reputation mechanism.}
    \label{fig:plot_types}
\end{figure}

We initially compare scenarios with and without our VMGuard anti-poisoning algorithm with respect to the type of the participating SIoT devices, as depicted in Fig.~\ref{fig:plot_types}. We observe a considerable reduction in accepting malicious SIoT devices to deliver their semantic data. Specifically, in the absence of our reputation mechanism, the VSP equally selects SIoT devices from various types. However, with the integration of our VMGuard mechanism, SIoT devices with positive reputations significantly dominate the selection for semantic data delivery (i.e., \textbf{type-0} SIoT devices as in Fig.~\ref{fig:plot_types}). Interestingly, we observe that malicious devices categorized as \textbf{type-2} exhibit a higher percentage than those classified as \textbf{type-1}. This behavior can be attributed to instances where \textbf{type-2} SIoT devices consistently opt for non-malicious behavior throughout certain time windows. This real-time behavioral observation allows the VSP to capitalize on these devices, gathering their semantic data. In other words, when SIoT devices from \textbf{type-2} choose not to attack in some time windows, our real-time mechanism detects this behavior and quickly update their reputation to allow collection of their semantic data. 

Another noteworthy strength of these results lies in their resilience to potential errors made by the reputation score calculation mechanism in scenarios where no actual attack occurs but some SIoT devices are flagged with negative scores. In other words, the system is capable to rectify false positives SIoT devices and allow their return to the market after few rounds which is notable.
This emphasizes the advantages of obtaining real-time feedback from VMUs.
Finally, one of the major advantages of this reputation technique is that we will have more free channels available to hire other SIoT devices with positive reputation scores. Otherwise, the loss would be at two levels, i.e., loss for channel payment and loss for tampered digital twin.




\subsubsection{Performance of The Reputation Memory Management}

\begin{figure*}[!h] 
     \centering
     \begin{subfigure}[b]{0.3\textwidth}
         \centering
         \includegraphics[width=\textwidth]{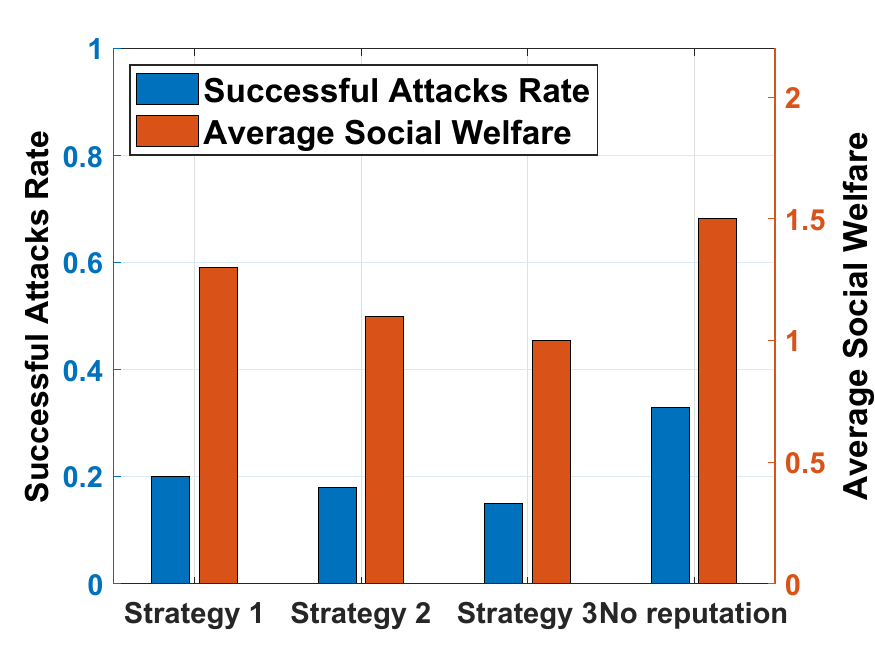}
         \caption{}
         \label{fig:memory_03}
     \end{subfigure}
     \hfill
     \begin{subfigure}[b]{0.3\textwidth}
         \centering
         \includegraphics[width=\textwidth]{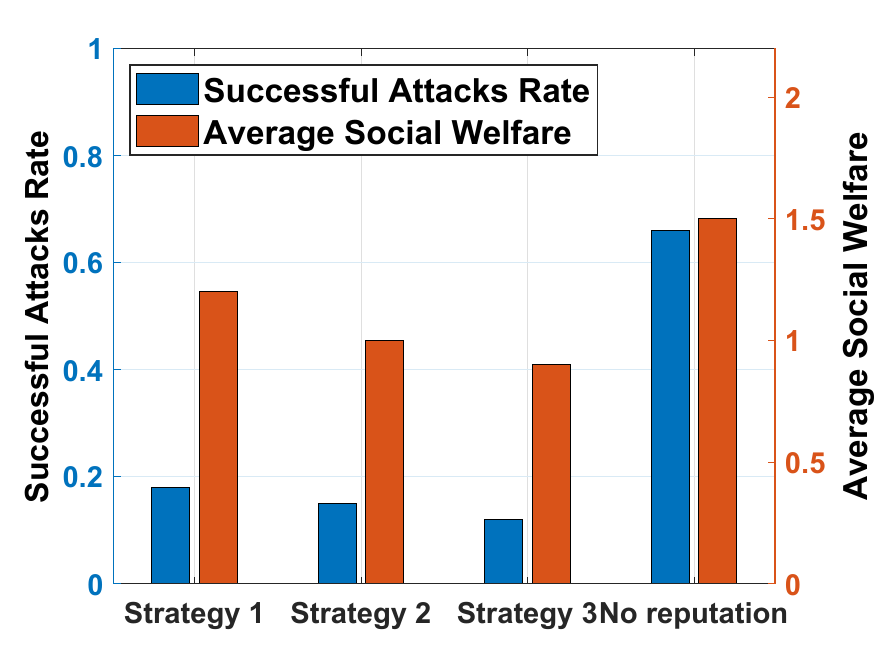}
         \caption{}
        \label{fig:memory_06}
     \end{subfigure}
     \hfill
     \begin{subfigure}[b]{0.3\textwidth}
         \centering
         \includegraphics[width=\textwidth]{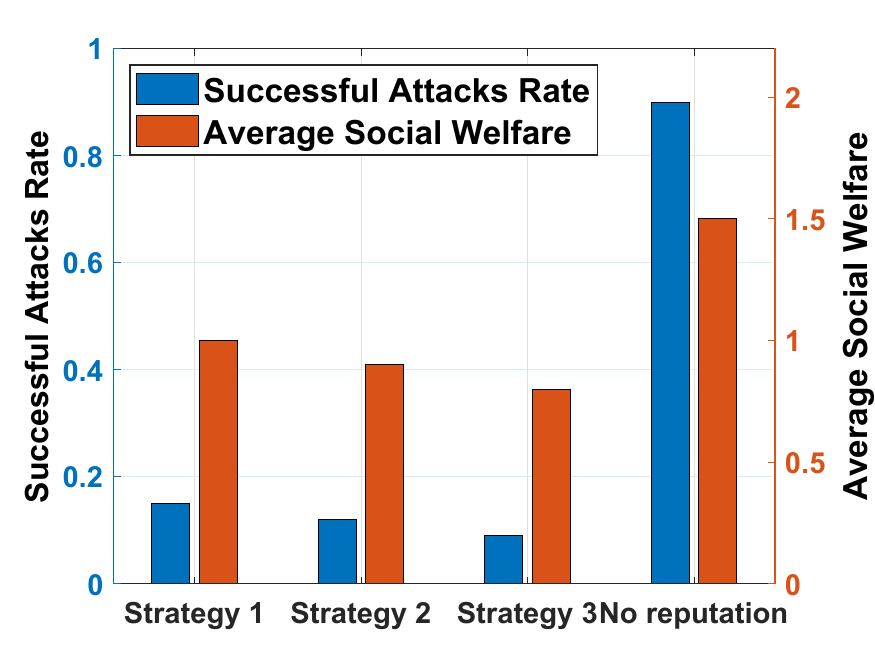}
         \caption{}
         \label{fig:memory_09}
     \end{subfigure}
        \caption{Average social welfare and successful attack rates in different scenarios. (a) probability of attack $P_{atk}=0.3$ (b) probability of attack $P_{atk}=0.6$ (c) probability of attack $P_{atk}=0.9$.}
        \label{fig:memory_management}
\end{figure*}

To evaluate the performance of best strategy of reputation memory management, we perform the following experiment. Here, we study the case of a single VSP interacting with a set of SIoT devices and VMUs. The number of participating SIoT devices and bids values are fixed for the entire times windows.
We also define the successful attacks rate metric as the ratio between successful attacks and all initiated attacks across all time windows. The experiment is conducted for an attacking SIoT devices from \textbf{type-1} under three different values of probability of attacks, i.e., $p_{atk}=0.3$, $p_{atk}=0.6$ and $p_{atk}=0.9$. We also set the number of the most recent interactions $D$ to $10$ for all the experiments where \textit{Strategy 2} and \textit{Strategy 3} are used.
Fig.~\ref{fig:memory_management} depicts the average social welfare of the system averaged across all time windows and the successful attack rate for three different strategies of handling the reputation database. Additionally, we show how these metrics vary in the case where no anti-poisoning mechanism is used (denoted as ``No reputation'').

First, we observe that the social welfare for the case where no anti-poisoning attack is used is the same for all values of the attack probability $p_{atk}$. This is argued by the fact that in the case where no anti-poisoning attack is used, all SIoT devices regardless of the reputation are accepted subject to their bids values. However, in the case where the reputation filtering is used, only SIoT devices with positive ratings are allowed to participate in the auction and then in the semantic data delivery, which are less in number compared to the full set of SIoT devices.
Additionally, we note that the average social welfare in the system without the reputation mechanism is higher than that achieved with our reputation-based mechanism. This outcome is consistent across all memory management strategies and is unexpected. Typically, a higher social welfare value indicates heightened system efficiency. However, this discrepancy is reasoned by our anti-poisoning mechanism, which does not directly tackle the maximization of social welfare. The social welfare captures system benefits based on the defined utility functions in~\eqref{eq:u_i} and~\eqref{eq:u_hat}, neglecting security performance degradation, notably digital twin tampering. This discrepancy becomes apparent when observing the significant high successful attack rate in scenarios without an anti-poisoning mechanism, as illustrated in Fig.~\ref{fig:memory_management}.

We also observe that \textit{Strategy 1} and \textit{Strategy 2} have similar performance for both successful attack rate and social welfare for different attack probabilities. Interestingly, \textit{Strategy 3} has slightly better performance than that of \textit{Strategy 1} and \textit{Strategy 2} in terms of reducing the successful attack rates. This is justified by the fact that \textit{Strategy 3} mixes both advantages of \textit{Strategy 1} and \textit{Strategy 2}, i.e., putting more weights to the most recent $D$ interactions. We also observe that as the attack probability $p_{atk}$ increases, successful attack rates decreases for the three strategies.
Therefore, our design incentivizes the attackers to reduce their attack frequency significantly in order for their attacks to have more effects, which brings great QoE to the VMUs.
By ensuring the integrity of the data used to construct the digital twin, we enhance the consistency and reliability of the services experienced by VMUs.

\subsubsection{Impact of Shared Public Profile Server}
\begin{figure}[!h] 
     \centering
     \begin{subfigure}[b]{0.49\textwidth}
         \centering
         \includegraphics[width=\textwidth]{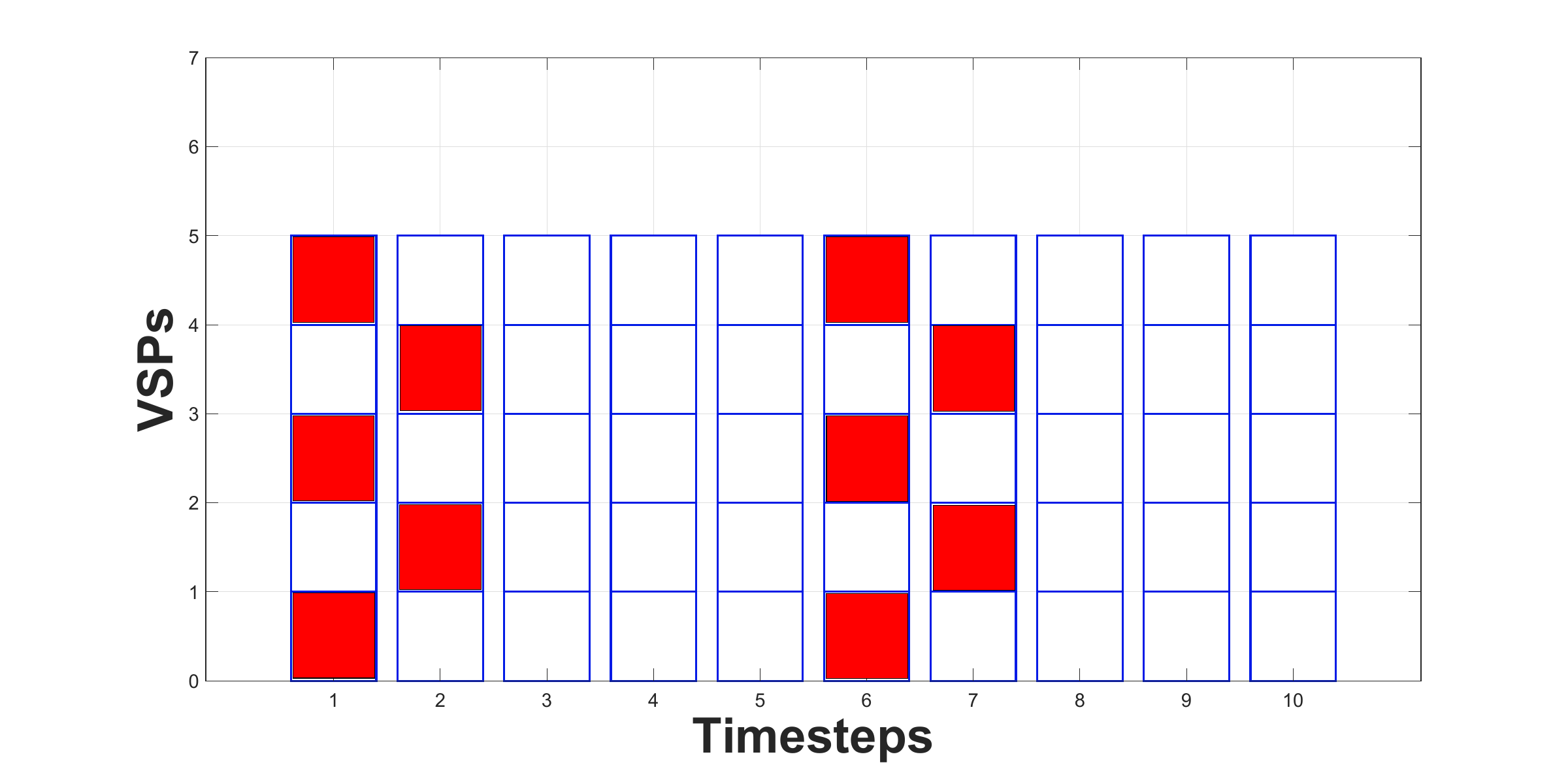}
         \caption{}
         \label{fig:non_cooperative}
     \end{subfigure}
     \hfill
     \begin{subfigure}[b]{0.49\textwidth}
         \centering
         \includegraphics[width=\textwidth]{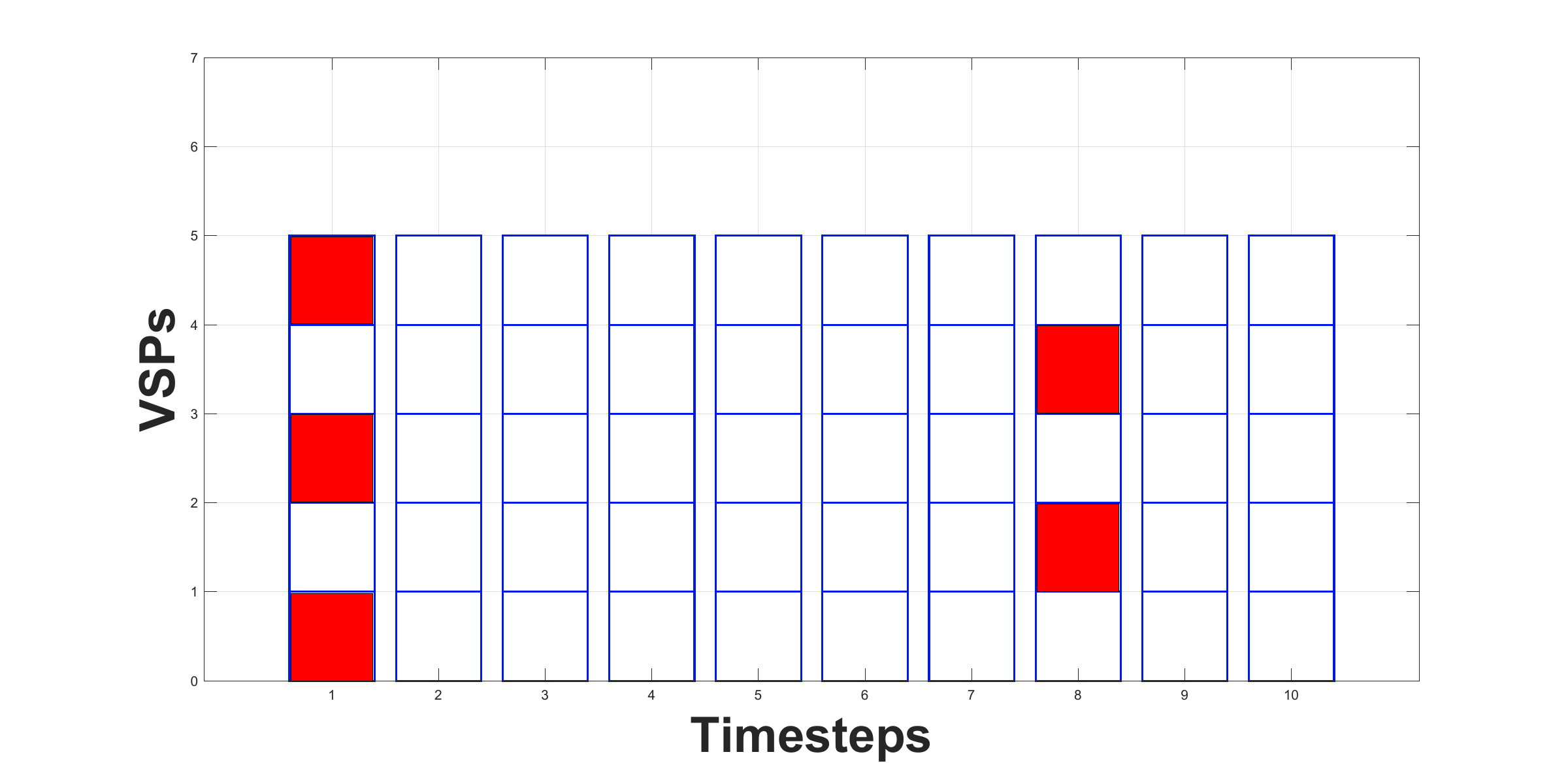}
         \caption{}
         \label{fig:cooperative}
     \end{subfigure}
     
        \caption{Successful poisoning attacks across timesteps on (a) non-cooperative VSPs design (b) cooperative VSPs design.}
        \label{fig:cooperative_vs_non}
\end{figure}

To show the impact of grouping the reputation scores of the SIoT devices in a public blockchain profile server, we perform the following experiment.
We follow the SIoT device with $ID=7$ which is considered to be a \textbf{type-2} attacker (i.e., alternate between attack and non-attack modes with $p_{atk}=0.5$) and has always a higher bid value to guarantee his selection in the auction algorithm. Additionally, for the better illustrative purpose, we consider the attacks on VSPs with pair IDs to be performed in pair timesteps while attacks on VSPs with impair IDs to be performed in impair timesteps. 
We also consider that we use the fading function based strategy for the reputation layer.
Fig.~\ref{fig:cooperative_vs_non} depicts the results of this experiment where 5 VSPs are considered to be either non-cooperative (i.e., do not use a shared profile server) or cooperative (i.e., use a shared profile server). In Fig.~\ref{fig:cooperative_vs_non}, boxes with red fills indicate the corresponding VSP has been successfully attacked.

We observe a significant reduction in successful attacks in the case of cooperative VSPs compared with the case of non-cooperative VSPs. Specifically, in timestep $t=1$ all attacks are successful for both cases. However, in the next timestep, attacks on cooperative VSP architecture are not successful even though the attacked VSPs in this timestep were not attacked previously while attacks on non-cooperative VSPs were successful. This shows a significant value of the effect of information sharing amongst VSPs in reducing poisoning attacks. 
In the subsequent timesteps, attacks are prevented for both cases due to the use of fading function based strategy for the reputation layer.
Interestingly, we also observe that attacks on non-cooperative VSPs architecture start occurring sooner compared to the case of cooperative VSPs.
We argue this behavior, which shows great advantage for our cooperative VSPs design, by the fact that the reputation scores in the cooperative design are calculated based on negative ratings received from several VSPs which brings more weight toward a final negative score as described in \eqref{eq:R_fin}. In the case of non-cooperative VSPs, as timesteps proceed further, the reputation score function will have only 1 negative rating which vanishes over time, pushing towards a positive rating score. This results shows the benefits of using both a fading function in the subjective logic model for the reputation calculation and of using the cooperative VSPs design (i.e., public profile server).






\subsection{Discussion}
To help better position our contribution, here we iterate the two main problems addressed in mechanism design and how they have been addressed: adverse selection and moral hazard.
Most of existing research has focused on the first issue, adverse selection, where the objective is to prevent a participant from taking advantage of asymmetric information that is not disclosed to all other participants. The standard formulation of these solutions consists of setting an ILP with the social welfare of the system to maximize subject to constraints refereed to as IC and IR.
The core idea here is to design a mechanism where no participant can deviate from the optimal solution without harming itself (i.e., decrease its utility which is also reflected in the decrease of the social welfare).
This direction is well investigated in the literature~\cite{Han_Niyato_Saad_Basar_2019, Niyato_Luong_Wang_Han_2020}. However, the second problem, moral hazard, is still not well studied. 

Moral hazard arises when a participant is incentivized to take risks or behave suboptimally or unethically because they do not fully bear the consequences of their actions. This issue is evident in the poisoning attack problem studied in this paper, where SIoT devices can retransmit the same data across consecutive time slots while still receiving payments. 
Since the social welfare maximization function does not account for the moral hazard problem, we cannot ensure that social welfare is maximized at all times. Consequently, our contribution is to strike a balance between maximizing social welfare and minimizing moral hazard behaviors (i.e., successful attack rate).

\section{Conclusion and Future Work}\label{section_conclusion}
Ensuring security in data collection is becoming crucial for virtual service providers gathering information through SIoT devices, as tampered content can compromise QoS/QoE. Addressing this challenge, our introduced incentive mechanism, VMGuard, employs a reputation-based strategy to dissuade malevolent behavior post-payment and incentivize SIoT devices to consistently engage truthfully in Metaverse services.
Our framework aims to prevent poisoning attacks in vehicular Metaverse systems. The strategy involves assigning reputation scores to SIoT devices based on VSP engagements, derived from VMUs' feedback via a subjective logic model. To retain "good" SIoT devices with false positive ratings, we implement a system that reduces the significance of past ratings, enabling the VSP to rely on current and accurate data.
Validation via extensive simulations demonstrates the efficacy of our approach. Our model effectively minimize poisoning attacks by malicious SIoT devices without unfairly excluding previously missclassified reliable devices from future market participation. The system's adaptability to address the dynamic nature of the environment and its ability to maintain market integrity without unjustly penalizing reliable devices stand as significant strengths. 
This framework offers a robust solution to enhance security in the vehicular Metaverse's data collection layer, ensuring integrity, and fostering trust among its diverse participants. 

As for the future work, it is of high significance to conduct real-world deployments and validation of the proposed mechanisms in actual vehicular Metaverse environments to assess scalability, practicality, and real-time effectiveness.
Additionally, we recommend integrating precise quantitative assessments through explicit QoS metrics like network latency and data accuracy, alongside QoE metrics such as user satisfaction scores, to provide a comprehensive and objective validation of our framework's improvements.
An important direction worth investigating is the robustness of our framework against coalition attacks where several malicious SIoT devices cooperate together to hinder the poisoning attack detection.
Finally, we should note that this work focused on vehicular Metaverse. Nevertheless, it is of great usefulness to explore its use cases for several crowdsourcing application where the participants can have malicious incentives to attack the system services.




\bibliographystyle{IEEEtran}
\bibliography{reference}

\end{document}